\documentclass[aps,prl,twocolumn, unsortedaddress,superscriptaddress,showpacs]{revtex4}

\usepackage{graphicx}
\newcommand{\Mt}{M$_\mathrm{J}$}
\newcommand{\Mts}{M$_\mathrm{J}$ }
\newcommand{\M}{\mathrm{M}_\mathrm{J}}
\newcommand{\J}{\mathrm{J}}
\newcommand{\K}{\mathrm{K}}

\newcommand{\ND}{ND$_3$\ }
\newcommand{\NH}{NH$_3$\ }
\newcommand{\NDn}{ND$_3$}
\newcommand{\NHn}{NH$_3$}
\newcommand{\cm}{cm$^{-1}$\ }
\newcommand{\cmn}{cm$^{-1}$}

\begin{document}

\title{State selective detection of velocity filtered ND$_3$ molecules}
\author{Benjamin Bertsche}
\affiliation{Fritz-Haber-Institut der Max-Planck-Gesellschaft, Faradayweg 4-6, D-14195 Berlin, Germany}
\author{Andreas Osterwalder}
\email{andreas.osterwalder@epfl.ch}
\affiliation{Ecole Polytechnique F\'ed\'erale de Lausanne, CH-1015 Lausanne, Switzerland}

\date{\today}

\begin{abstract}
Translationally cold and slow \ND is prepared by filtering the slow molecules from a thermal gas-phase sample using a curved electrostatic hexapole guide.
This filter, like the curved quadrupole guide introduced by Rempe et al. in 2003\cite{Rangwala:2003p4185}, selects molecules by their forward velocity and effective electric dipole moment.
Here we describe two main modifications with respect to previous work: 1. A hexapole guide is used instead of a quadrupole, thus producing a harmonic potential for the linearly Stark-shifted levels of \NDn. 
The curved guide is combined with a straight hexapole guide with independent high-voltage supplies to allow for band-pass velocity filtering.
2. State-selective laser ionization is used to obtain time- and state selective detection of the guided molecules.
This enables the experimental determination of the rotational state population of the guided molecules.
\end{abstract}


\maketitle
\section{Introduction}
Cold neutral molecules provide many intriguing possibilities in precision spectroscopy and chemical dynamics.\cite{Krems:2009p3148,Bell:2009p4182}
Many methods have been developed over the past years to prepare neutral molecules in the millikelvin to microkelvin range.
These methods have been applied in high-resolution spectroscopy \cite{vanVeldhoven:2002p2056,hudson:06} as well as collision studies.\cite{Gilijamse:2006p2988,Willitsch:2008p4113,Knoop:2010p4763}
Two main approaches are currently available for the production of cold neutral molecules (see \cite{Krems:2009p3148} and references therein): 
1. They can be assembled from cold atoms using lasers or electric fields.
2. Molecules with a permanent electric or magnetic dipole moment can be prepared at high laboratory frame-of-reference velocities and then be decelerated using time-varying electric, magnetic, or optical fields \cite{Osterwalder:2010p4866,vandeMeerakker:2008p4104,Hogan:2008p2965,Fulton:2006p4875}.
Electric fields have also been used to guide polar molecules along straight and curved trajectories, and it has been shown that an electrostatic guide can be bent into a circle to obtain a storage ring \cite{Crompvoets:2005p4878}.
In those experiments, an hexapole guide was used, very similar to the one employed in the present study.
In 2003, Rempe and co-workers demonstrated a relatively simple technique to obtain translationally cold polar molecules by filtering the slow ones from a thermal sample using an electrostatic quadrupole guide\cite{Rangwala:2003p4185,Junglen:2004p4126,Sommer:2009p3065,Tarbutt:2009p4094}.
In a thermalized sample, even at room temperature there always exists a small fraction that moves at very low velocities.
Isolation of these molecules from the sample provides a continuous flux of translationally cold molecules.
Since the initial flux from the source can be considerable, even this small fraction corresponds to densities that are comparable to those obtained from other methods.
Rempe et al. \cite{Rangwala:2003p4185} first demonstrated this approach by coupling a continuous effusive source of ammonia with a curved electrostatic quadrupole guide.
Since then, the method has been applied to several other molecules, and has been used, e.g., for detailed studies of ion-molecule reactions \cite{Willitsch:2008p4184,Willitsch:2008p4113,Bell:2009p4111}.

Inside an electrostatic guide the inhomogeneous electric field pushes molecules in certain (low-field seeking) quantum states toward the center of the guide.
In a bent guide, the centrifugal force is added to the trapping force, and those molecules where the centrifugal force exceeds the trapping force are ejected from the guide.
The velocity distribution of the guided molecules can be determined by measuring the time-of-flight (TOF) distribution at the end of the guide.
Typically, this was done by coupling the guide to a mass spectrometer where the molecules were ionized by electron bombardment.
This provides accurate measures of flight times and of the flux as a function of time.
However, it does not reveal any information about the internal state population.
The selection in an electrostatic guide is based on the forward velocity and the effective dipole moment.
Since there is no cooling or deceleration, the selectivity both in terms of velocity and internal levels is based on these two handles alone.
The effective dipole moment depends on the permanent dipole moment of the molecule and on the rotational level.
As a consequence, the rotational temperature of the guided molecules will not be the same as that of the molecules before the filter.
To what extent it is changed, however, has been a matter of theoretical predictions so far, because no state-selective detection was used.
Calculations \cite{Bell:2009p4111} predict a moderate change of the internal temperature, and the production of rotationally cold velocity-filtered molecules has only been achieved by the coupling of a guide and a buffer-gas cooled source.\cite{VanBuuren:2009p4123,Sommer:2009p3065,Patterson:2007p4868}

In the present study we combine a segmented, partially curved, electrostatic hexapole guide with resonance-enhanced-multiphoton ionization (REMPI) for time- and state-resolved detection of the guided molecules.
This provides the following advantages:
1. The hexapole guide is split into a straight portion and a curved portion.
The curved section has a large radius of curvature (125 mm vs. ca. 15 mm in previous studies).
The two sections are connected to separate high-voltage switches to implement a velocity band-pass filter.
In a similar experiment, Rempe et al. have recently demonstrated the selection of molecules within a $\approx$ 5 m/s wide band \cite{Sommer:2010p4769}.
Repeated application of such high-voltage pulses can be used to produce a quasi-continuous flow of molecules within a selected velocity range.
2. The large radius of curvature provides very well-defined conditions and enables the accurate calculation of particle trajectories through the structure.
This is used here to characterize the device and to extract guiding probabilities for different rotational states of the molecule.
3. Using a hexapole instead of a quadrupole guide is advantageous for states which are shifted linearly in an electric field.
A hexapole guide possesses a nearly harmonic electric field transversely, which in turn corresponds to a harmonic potential for a linearly shifted state.
4. Using state-selective REMPI to detect the guided molecules enables the direct characterization of the rotational temperature.

Ammonia was used in all experiments presented here.
As a prototypical symmetric top molecule it has been studied extensively by different types of spectroscopy, including in particular the sensitive state-selective detection of \NH and its isotopomers using (2+1) REMPI\cite{ASHFOLD:1988p4430}.
A vast number of studies is available that measure in detail the structure of the $\tilde{X}$ ground state as well as of the excited $\tilde{B}$ state which was also used in the present study.
This state is conveniently excited using two UV photons around 315 nm, and a third photon of the same wave length efficiently ionizes the molecules.

\begin{figure}
\includegraphics[width=3in]{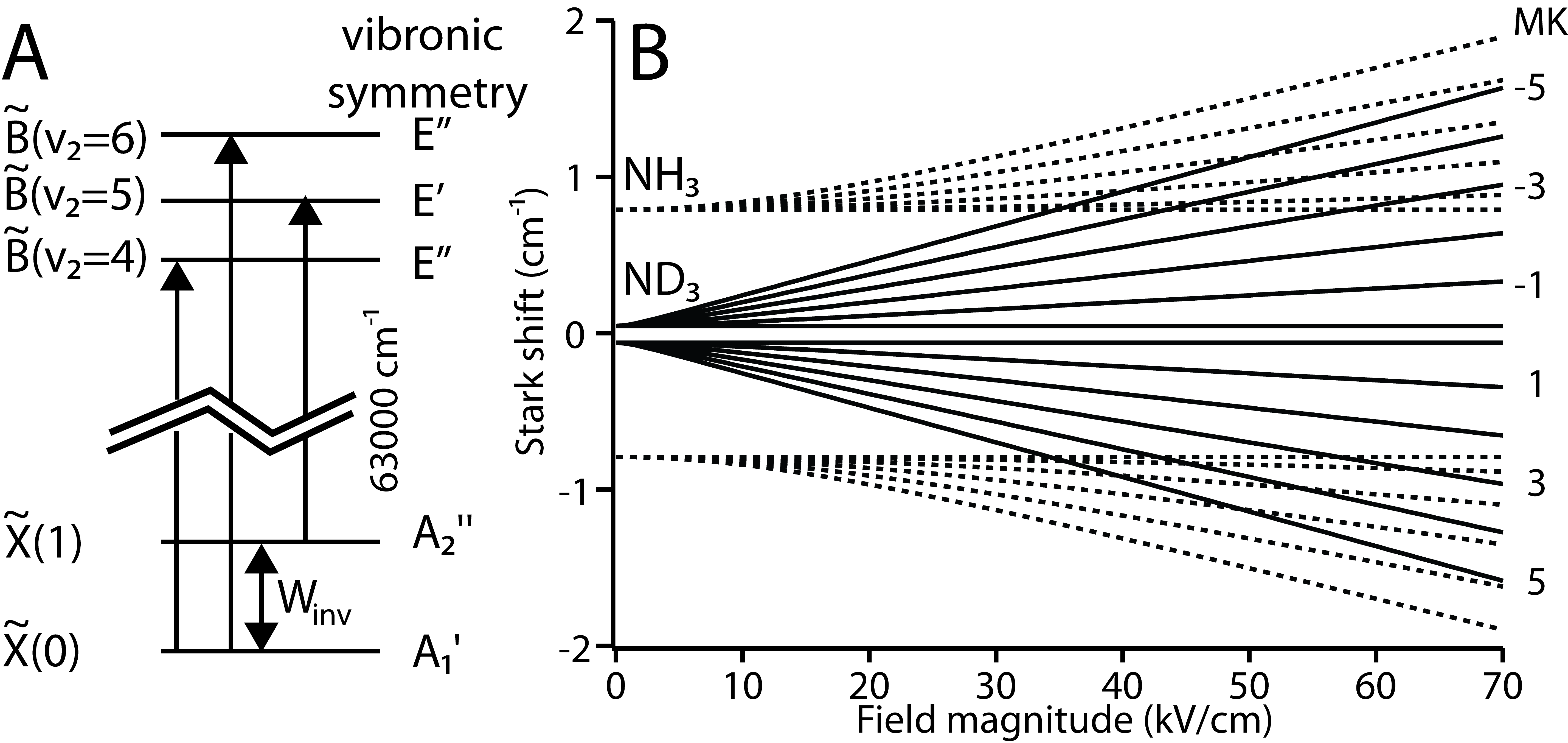}
\caption{\label{fig1}A) level scheme of ammonia with symmetry labels for \NDn . The inversion splitting in the ground state, $W_{inv}$ is indicated, and the arrows show symmetry-allowed transitions. B) Stark energies for levels with $J=5$ for \ND (solid lines) and \NH (dashed lines).}
 \end{figure}

The ground state of ammonia has a double minimum potential, originating from the umbrella motion where the N atom is moved through the plane containing the H atoms.
The two pyramidal minima on the potential energy surface are connected via an energy barrier ($\approx$2000 \cm high in \NHn) which leads to a tunneling splitting of the ground state into two levels with opposite symmetry.
These two levels are generally labeled as $\tilde{X}(0)$ and  $\tilde{X}(1)$ and are energetically split by the inversion splitting $W_{inv}$ (see fig. \ref{fig1}A).
Inspection of the symmetries of the $\tilde{X}$ and $\tilde{B}$ states, including the umbrella vibrational mode $v_2$ in the excited state produces the following selection rules\cite{Ashfold:1985p4863}:
\begin{eqnarray*}
\tilde{B}(v_2=\mathrm{odd})\not\leftarrow\tilde{X}(0),& &\tilde{B}(v_2=\mathrm{odd})\leftarrow\tilde{X}(1), \\
\tilde{B}(v_2=\mathrm{even})\leftarrow\tilde{X}(0), &\mathrm{and\ } &\tilde{B}(v_2=\mathrm{even})\not\leftarrow\tilde{X}(1). 
\end{eqnarray*}

The energetic Stark-shift of an ammonia-molecule in an electric field with magnitude $E$ is given, for a particular rotational state $\langle \J,\K|$, by \cite{Crompvoets:2005p4878}
\begin{equation}\label{starkshift}
\mathrm{W(J,K,M)}=\pm\left(\frac{W_{inv}}{2}+\sqrt{\left(\frac{W_{inv}}{2}\right)^2+\left(-\mu E\kappa\right)^2}\right),
\end{equation}
where $\kappa=\frac{\M\K}{\mathrm{J(J+1)}}$, $\mu$ is the permanent electric dipole moment ($\mu=1.47$ D for \NDn), and the positive (negative) energies correspond to the upper (lower) component of the inversion doublet.
\Mts is the projection of J on the field axis ($-\mathrm{J} > \mathrm{M}_\mathrm{J} > \mathrm{J}$), $W_{inv}$ is the inversion splitting ($W_{inv}$(\NDn)=0.05 \cm; $W_{inv}$(\NHn)=0.79 \cmn).
Levels from the upper component (with $|\mathrm{M}_\mathrm{J}|>0$) are shifted to higher energies in increasing electric field magnitude (low-field-seeking states; lfs) while those from $\tilde{X}(0)$ are shifted to lower energies (high-field-seeking states; hfs).

The large inversion splitting in normal ammonia leads to a quadratic stark shift of all levels up to several 10 kV/cm, while in deuterated ammonia the shift is linear already at very low fields (see fig.\ref{fig1}B).
In an inhomogeneous electric field, lfs levels feel a force toward regions with lower field magnitude.
In a quadrupole guide, the field increases linearly with increasing radius, thus producing harmonic potential for \NHn , but a linear one for \NDn .
Instead, a hexapole creates a harmonic potential for \NDn .
In all measurements described here only \ND has been used.
Fig. \ref{fig1}B) shows the energy levels for the (J=5, K=5) components of the \ND ground vibronic state in electric fields between 0 and 70 kV/cm.
At zero field, levels with different \Mts are degenerate, and the two remaining levels are split by $W_{inv}$. 
With increasing field, the components split and are shifted to higher or lower energy.

An electrostatic hexapole, constructed by arranging six electrodes on the corners of a hexagon and applying alternating positive and negative voltages, provides a 2-dimensional trap.
The resulting electric field (simulated using finite element methods \cite{Anonymous:2009p4864}) is plotted in fig.\ \ref{expt}B) where - for the case of $\pm$8 kV on the electrodes -- the innermost contour line lies at 10 kV/cm, and the spacing between lines is also 10 kV/cm.
At small radii the field is to a good approximation cylindrically symmetric. 
A cross-section through the field along the horizontal line shown in fig.\ \ref{expt}B) is plotted in fig.\ \ref{expt}C). 
The field increases harmonically with increasing radius as is confirmed by the overlaid parabola in the figure.
The force is purely transverse, and in a straight guide, lfs-molecules are guided irrespective of their forward velocity.
Only molecules in lfs states are guided since hfs states are pushed away from the center of the guide and lost.
In a curved guide the molecules feel a centrifugal force in addition to the guiding force, and they are guided as long as the latter is larger than the former.
Equating the centrifugal force 
\[F_C=\frac{mv_l^2}{R+r_0}\]
and the opposing, trapping force
\[F_S=-2\mu E\kappa\frac{r}{r_0^2}\]
defines a maximum guidable forward velocity
\begin{equation}\label{vmax}
v_{l,max}=\sqrt{\frac{2\mu E_{max}}{m}\kappa\left(1+\frac{R}{r_0}\right)}.
\end{equation}
Here, $R$ is the radius of curvature of the guide, $r_0$ is the inner radius of the guide, and $r$ is the radial position of the particle.
Eq.\ref{vmax} was obtained by assuming a harmonic potential and $W_{inv}=0$.
Since both these assumptions slightly improve the guiding capacity for a particular molecule, eq.\ref{vmax} is an overestimate of the actual value.
For a complete description of the dynamics of ammonia molecules in a curved hexapole guide we refer to reference \cite{Crompvoets:2005p4878}. 

\section{Experimental}
\begin{figure}
\includegraphics[width=3in]{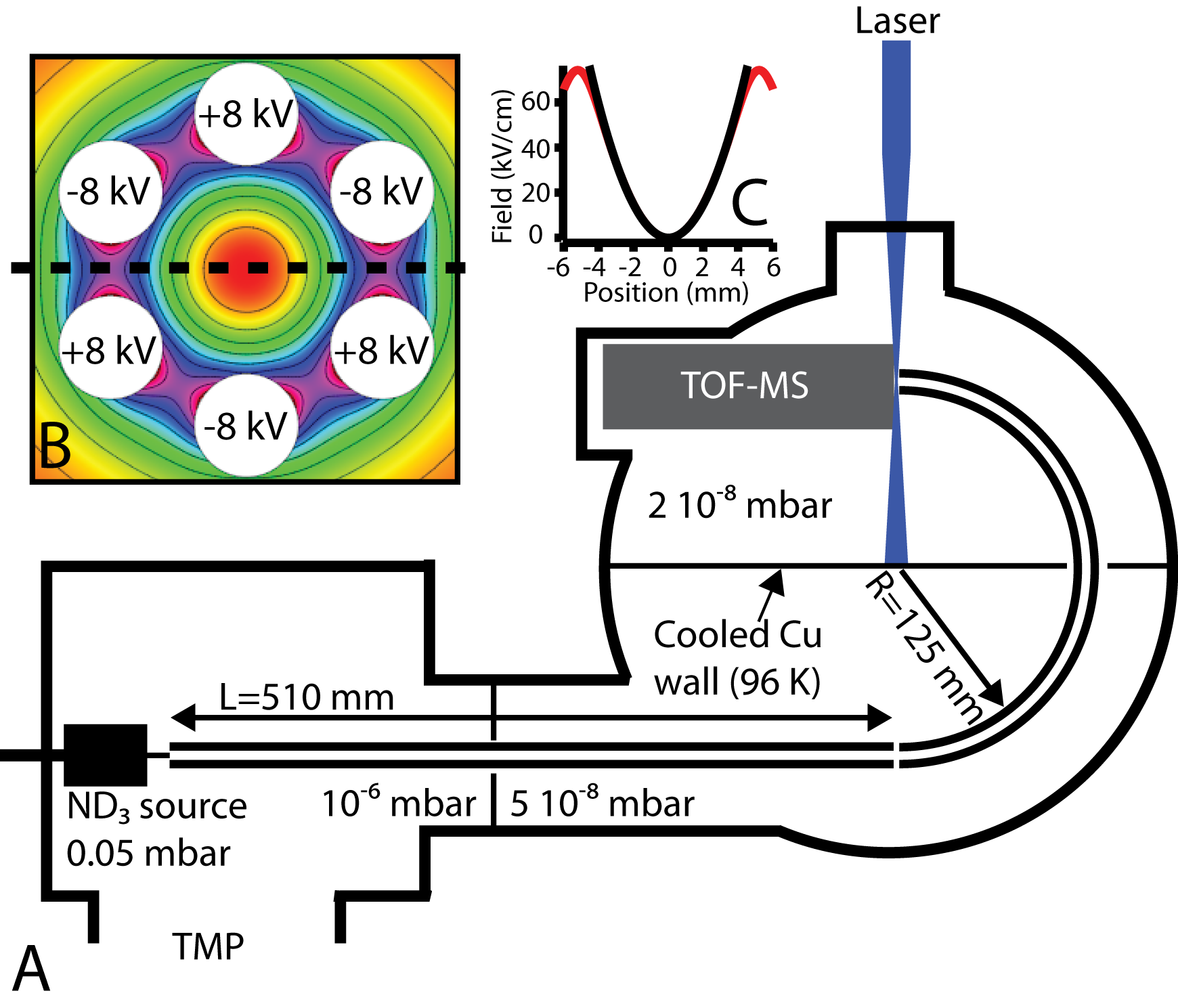}
\caption{\label{expt}A) Sketch of the experimental setup. B) Cross-section through a hexapole guide as used in the present experiments. The white circles are the positions of the electrodes with voltages applied as indicated. Contour lines start at 10 kV/cm (at the center) and are spaced by 10 kV/cm. C) Red line: cut through the potential shown in panel B), black line: parabola to fit the red line. }
 \end{figure}
The experimental setup is sketched in fig.\ \ref{expt}A).
A continuous beam of \ND is generated from an effusive source (left side of the figure), built from a cylindrical copper container (20 mm inner diameter) with a ceramic outlet tube (5 mm long, 2 mm inner diameter).
The pressure inside the source is ca. 5$\cdot$10$^{-2}$ mbar during operation.
The tip of the ceramic tube is at the beginning of the guide, and the \NDn-beam at ca. 300 K directly enters the electrostatic hexapole guide.
The hexapole guide consists of a straight (510 mm long) and a curved segment (a half-circle with radius of curvature 125 mm) where the former connects two differentially pumped high vacuum chambers.
Both segments are built from six individual polished cylindrical stainless-steel electrodes arranged on a 6 mm diameter circle, as shown in fig.\ \ref{expt}B).
The electrode diameter is 4 mm, leaving a circular open space in the center with a radius of 4 mm.
In the present study, the hexapole voltages were varied between $\pm$8 kV and $\pm$2 kV on the curved guide, while the straight guide was operated by keeping three electrodes at ground potential while the other three were always at +6 kV.
Three fast (Behlke; ca. 200 ns rise time) high-voltage switches were used to rapidly switch the two guide segments on and off individually to record time-of-flight (TOF) traces at the end of the curved guide.

For the REMPI-detection of \NDn, the frequency-doubled output of a pulsed tunable dye laser (Fine Adjustment Pulsare; ca. 15 mJ/pulse at 315 nm), pumped by the second harmonic of a pulsed Nd:YAG laser (Innolas Spitlight 1000; 20 Hz, 10 ns pulse length) is focused, using a 500 mm cylindrical lens, between the first and second plate of a Wiley-McLaren-type TOF mass spectrometer (MS).
The MS is aligned colinearly with the end of the curved hexapole guide, and the distance from the end of the guide to the intersection with the laser beam is ca. 15 mm.
The first extraction plate is positioned directly behind the guide and has a 20 mm hole, covered by high transmission Ni mesh to shield the mass spectrometer from the guide-HV.
To further reduce possible perturbations, the guide is switched off 50 $\mu$s before the laser is fired.
During this time, the fastest molecules move $\approx$ 10 mm which is less than the distance between guide and laser beam.
The molecules are ionized with the dc-extraction field on, and ions are accelerated to 2.5 keV and detected on an microchannel plate detector.

The differentially pumped high-vacuum chamber is split into three sections and pumped by two turbomolecular pumps.
A 20 mm opening between the source\--chamber (pumped by a Pfeiffer TMU1600; pumping speed 1400 l/s) and the second chamber (Pfeiffer TMU1400; 980 l/s) reduces the gas flux, while leaving enough space for the straight hexapole to pass through.
A copper plate, that can be cooled to ca. 90 K using liquid nitrogen, is installed in the second chamber.
It serves to further reduce the gas flux into the detection region, and at the same time cryogenically reduces the overall \ND background.
During operation, the pressure in the first chamber rises to 3$\cdot$10$^{-6}$ mbar, while in the detector chamber it rises from $<$5$\cdot$10$^{-9}$ to 2$\cdot$10$^{-8}$ mbar.
A considerable background ion signal is observed upon REMPI-detection even when no HV are applied to the guides.
Under normal operating conditions for the pre\-sent type of TOF-MS, this background signal is of similar amplitude as the desired signal from the guided molecules.
In order to discriminate against it, the electric field in the first section of the MS is reduced to ca. 10 V/cm.
This is still sufficient to extract all ions from the guided molecules, which have an anisotropic velocity distribution peaked towards the detector.
Because only one mass has to be detected from molecules that are generated in a very small volume, the peak in the TOF spectrum remains sufficiently sharp.
On the other hand, the thermalized background gas is not efficiently extracted, and the signal from these molecules is strongly broadened and reduced in intensity.
This way, the signal/background ratio was improved to $>$100 in the present experiment.

\section{Results}
\begin{figure}
 \includegraphics[width=3in]{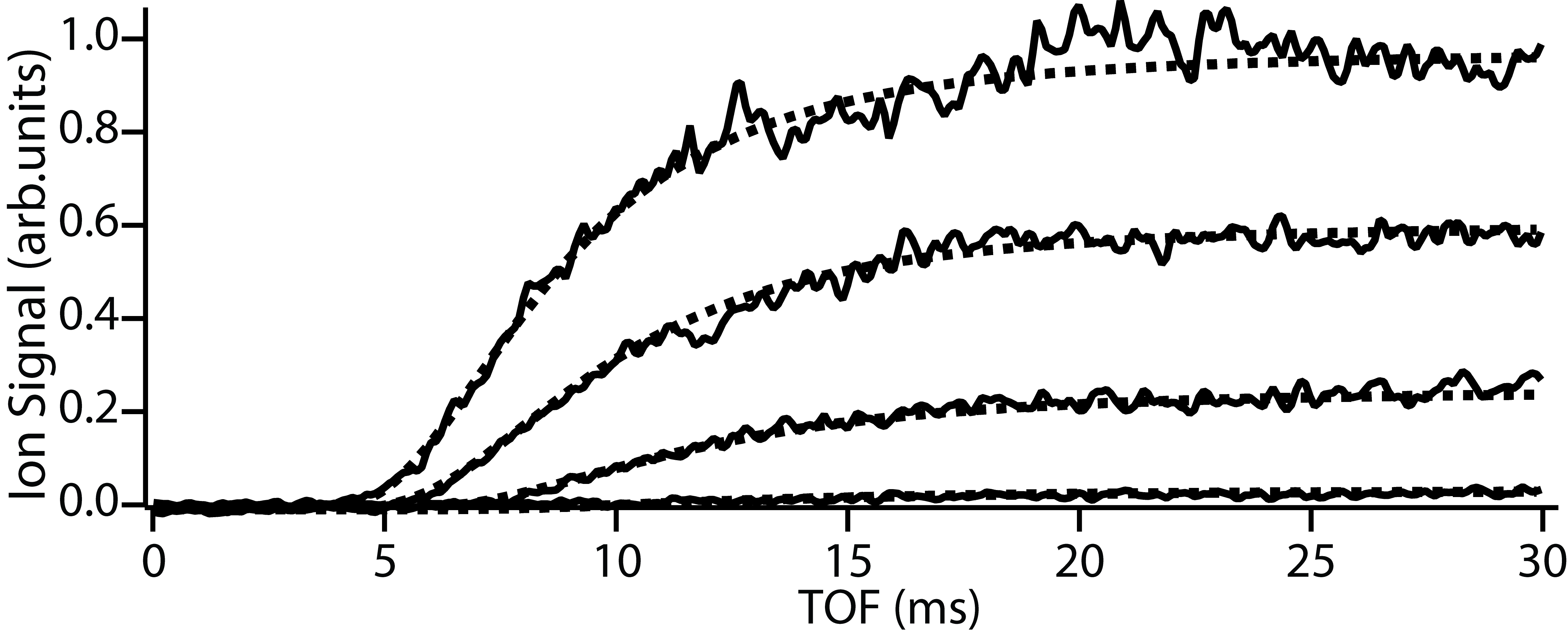}
\caption{\label{comp}Solid curves: experimental TOF-traces for guided molecules using $\pm$8, $\pm$6, $\pm$4, and $\pm$2 kV on the curved guide. These traces represent the signal from a single REMPI-transition. Dashed curves: analytic calculation assuming translational temperatures of 9 K, 7 K, 5 K, and 4 K, respectively. }
 \end{figure}
Time-of-flight traces from measurements using different voltages on the curved guide are shown as solid curves in fig.\ \ref{comp}.
They were recorded with -- from top to bottom -- $\pm$8, $\pm$6, $\pm$4, and $\pm$2 kV on the curved guide, respectively.
Each trace was recorded by switching the two guide segments on at the same time, and scanning the delay time between the application of the HV and the laser pulse.
The laser wave length was set to a particular transition in the $\tilde{B}(v_2'=5)\leftarrow\tilde{X}(1)$ band, and all traces shown here were recorded by monitoring only this single transition.
The $\pm$8 kV-signal starts rising at $t_{onset}\approx$5 ms, corresponding to $\approx$200 m/s in approximate accordance with eq.\ref{vmax} using $E_{max}$=60 kV/cm and an average value for $\kappa$, $\bar{\kappa}$=0.5.
The signal rises quickly and converges to a maximum which it reaches at around 15 ms.
The signal increase reflects the additional components of the velocity distribution, which is contained in the derivative of the TOF-traces.
Reducing the guide-voltage has three effects: 1. The onset is shifted to later times, 2. the slope is reduced, and 3. the final signal level is lower.
Eq.\ref{vmax} predicts $t_{onset}(V_{guide})\propto\sqrt{\frac{1}{V_{guide}}}$, corresponding to $t_{onset}(V_{6 kV})=5.7$ ms, $t_{onset}(V_{4 kV})=7.1$ ms, and $t_{onset}(V_{2 kV})=10$ ms, in accordance with observations.
Since the signal is proportional to the derivative of the velocity distribution, the shifted position of the turning point at lower voltages confirms that the most probable velocity is lower at lower velocities.
Finally, the lower convergence value implies an overall reduction of the flux, which is consistent with the reduced threshold velocity.

Two complementary approaches were chosen to extract translational and internal temperature information from the experimental data: 
1. Fitting of an analytic function, and 2. exact trajectory calculations:
As is described in\cite{Junglen:2004p4126}, the velocity distribution at the end of the filter is still described by a Maxwell-Boltzmann distribution. 
Because in the experiments described there, the detection procedure measured the flux, the distribution had to be scaled by a factor $1/v$.
Since in the present study the density is probed, this scaling factor is redundant and the correct description is a 3D Maxwell-Boltzmann distribution:
\begin{equation}\label{maxBol}
\rho(v)=Nv^2\exp{\left(-\frac{W_{kin}(v)}{kT}\right)},
\end{equation}
where $N=\sqrt{\frac{2}{\pi}}\left(\frac{m}{kT}\right)^{\frac{2}{3}}$ is a normalization constant, $k$ is the Boltzmann constant, and $T$ the temperature.
A TOF-distribution is obtained by integrating eq.\ \ref{maxBol} over time:
\begin{equation}\label{TOFint}
\rho(t^\prime=TOF)=\int_{v^\prime=\frac{\ell}{t^\prime}}^\infty\rho(v=\frac{\ell}{t})dv,
\end{equation}
where $\ell$ is the total flight distance.
Fitting eq.\ \ref{TOFint} to the experimental data directly provides the temperature of the guided molecules.
The dashed curves in fig.\ref{comp} show the TOF traces obtained for  temperatures of 9 K, 7 K, 5 K, and 4 K, respectively (from top to bottom), that very well fit the corresponding experimental data.
To determine the rotational temperature, the TOF traces are simulated by computing complete particle-trajectories through the electrostatic structure.
The two calculation methods are complementary: trajectory simulations do not directly yield a final translational temperature, while information on individual trajectories, or on the rotational temperature are not accessible via the semi-analytic method.
Results from trajectory calculations are presented later in this article to describe the different guiding probabilities for different rotational states.

\subsection{Velocity bandpass-filter}
\begin{figure}
\includegraphics[width=3.4in]{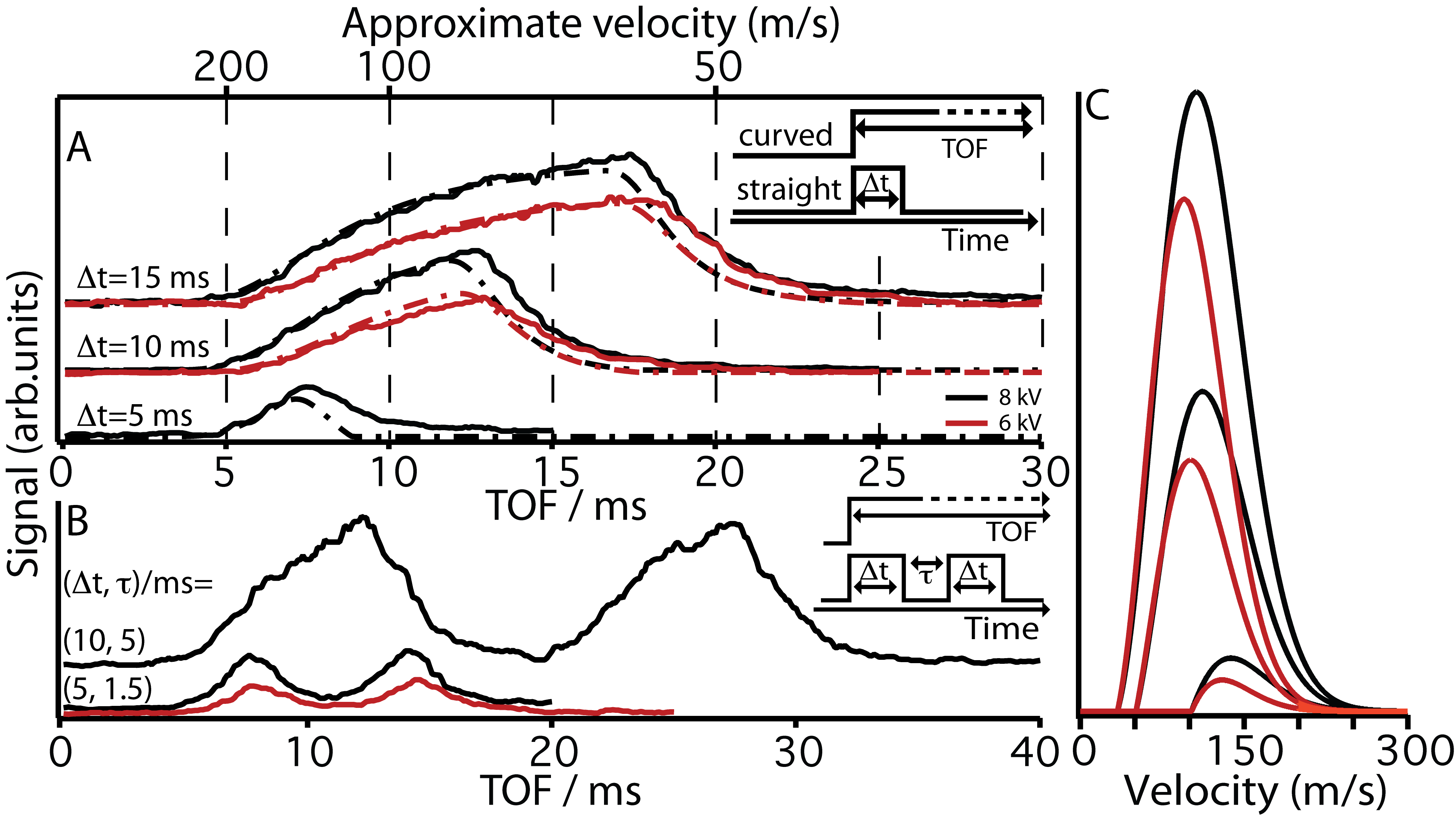}
\caption{\label{switched}A) TOF traces for bandpass-filtered molecules using 8 kV (black lines) and 6 kV (red lines). Solid lines are experimental data, dash-dotted lines simulations. The straight guide was kept on for a limited time only such that molecules below a certain threshold velocity were not detected. B) Multi-pulse sequence for a train of two pulses with filtered velocities at different voltages and switching sequences. C) Velocity distributions corresponding to the simulated traces in panel A). Black (red) show the 8 kV (6 kV) results. Values for $\Delta t$ are indicated in panel A).}
 \end{figure}
The measurements from the previous section enable the implementation of an upper velocity threshold via the guiding voltage.
Filtering against slow molecules is not possible in a static hexapole, but it is possible by switching the HV on the first, straight segment.
Figure \ref{switched}A)  shows TOF traces recorded at $\pm$8 kV (black lines) and $\pm$6 kV (red lines), and using the switching scheme sketched on the right side of the panel.
The TOF traces were recorded by switching both guides on at the same time, but switching the straight guide off after a time $\Delta$t.
The timing of the laser pulse was delayed relative to the on-time of both guides such that the TOF represents the flight-time through the entire guide, and the laser wave length was parked on the same transition as for fig. \ref{expt}.
Switching the straight guide off at time=$\Delta t$ means that molecules with a velocity less than $v_{min}=\frac{510 \mathrm{\ mm}}{\Delta t}$ will not be guided into the curved guide, and will thus not be detected.
The top, middle and lower traces in fig. \ref{switched}A) were recorded using $\Delta$t=15 ms, 10 ms and 5 ms, respectively, corresponding to $v_{min}\approx$ 35 m/s, 50 m/s and 100 m/s.
Upon switching off the straight guide, the signal decays at a rate determined by the velocity composition of the molecules inside the curved guide at TOF=$\Delta$t, but also by the time required to purge the straight guide.
The signal-reduction in the TOF-traces shown in fig. \ref{switched}A) does not start at TOF=$\Delta$t but later.
Until the straight guide is turned off, the flux into the curved guide is unchanged.
At TOF=$\Delta$t, the curved guide is not immediately emptied but merely does the influx cease.
Thus, the signal measured in the few ms after switching off the straight guide is determined by the velocity distribution of the molecules that are still inside the curved guide.
Only after the last of the fastest molecules have traversed the entire curved guide does the signal start to drop.
In the case of $\pm$8 kV, this time is $\approx\frac{390 \mathrm{\ mm}}{200 \mathrm{\ m/s}}$=1.8 ms, in accordance with experimental observation.

Semi-analytic simulations of these measurements are shown as dash-dotted curves.
Overall, good agreement is observed at short times-of-flight.
Important discrepancies, however, are present at the falling edges of the signals.
These  differences are explained by the dynamics that lead to the draining of the straight guide upon switching off of the voltages, as described below.
The velocity distributions resulting from these switching schemes are shown in panel C) of fig. \ref{switched}.
Black (red) curves show the distributions for 8 kV (6 kV) on the curved electrodes, with switching times of 15 ms, 10 ms, and 5 ms, respectively on the straight guide, for the curves from top to bottom.

A quasi-continuous beam of velocity-selected molecules is obtained by repeatedly switching the straight guide on and off.
Examples for such measurements are shown in fig.\ \ref{switched}B.
The pulse sequence is sketched on the right side of the panel, and the three traces are for ($\Delta$t=10,$\tau$=5) ms at $\pm$8 kV (top solid trace), and at ($\Delta$t=5,$\tau$=1.5) ms (bottom traces) at $\pm$8 kV (black trace) and $\pm$6 kV (red trace), respectively.
Here, $\Delta$t and $\tau$ are the pulse length and delay of the HV pulses, respectively.
In each case, two packets were recorded, but the scheme could easily be extended to an infinite stream of packets.
The maximum duty cycle that can be achieved is given by the time required for all molecules to be ejected from the straight guide.

\begin{figure}
\includegraphics[width=3in]{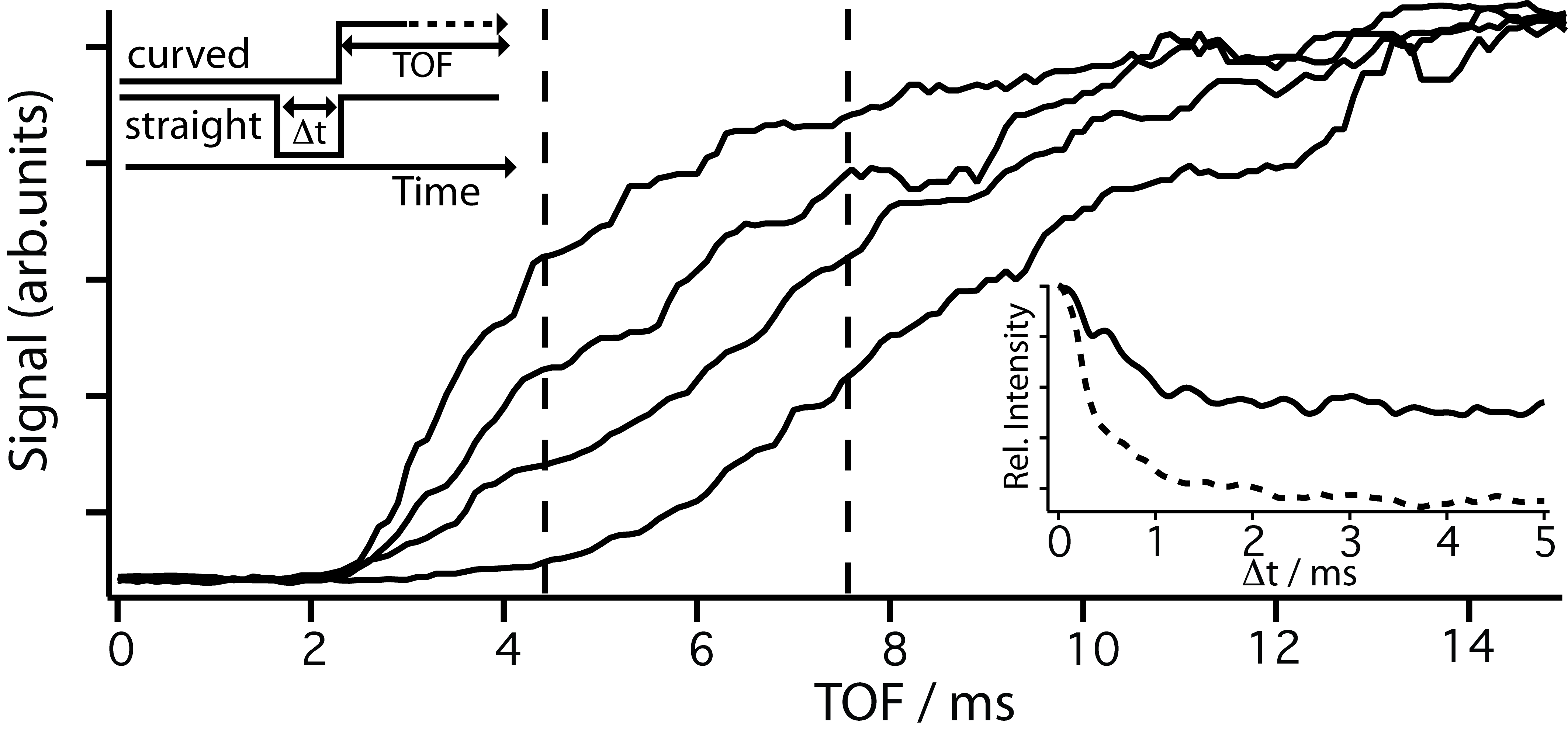}
\caption{\label{decay}Signal-decay upon switching the straight guide off. Main Panel: TOF traces or different time delays $\Delta$t. Top to bottom: $\Delta$t=10$\mu$s, 0.25 ms, 0.5 ms, and 5 ms. Inset: Signal-decay in the guide as a function of $\Delta$t, recorded at total TOF=4.5 ms (dashed line) and 7.5 ms (solid line).}
 \end{figure}

To types of experiments were done to obtain further information on this process:
The main panel in fig. \ref{decay} shows measurements where the straight segment was off only for a time $\Delta$t, and both guide segments were switched on simultaneously (as sketched in the top-left of the figure).
Traces shown here were recorded with $\Delta$t=0.01 ms, 0.25 ms, 0.5 ms, and 5 ms (top to bottom), where in each case the laser-delay was scanned to record the flight time.
In such an experiment, the straight guide is thus turned on most of the time and guides molecules to the entrance of the curved segment.
Upon switching the straight guide off, the molecules are not confined anymore and can leave.
When the guide is turned back on, some particles will be trapped again, and they will continue to fly through both sections and be detected.
Two extreme cases are observed for $\Delta t$ very short ($\Delta t=10 \mu s$) or very long ($\Delta t=5$ ms).
In the first case, the straight guide is switched off for a time to short for a significant portion of the molecules to leave it.
As a consequence, the TOF-trace is similar to one where the straight guide had been on all the time, and the observed flight times correspond to a distance of $\approx$39 cm, the curved guide alone.
In contrast, when the straight guide is switched off for a sufficiently long time for all molecules to leave it, a TOF distribution is observed that corresponds to flight through the entire (straight+curved) guide with a total distance of $\approx$90 cm.
This case is the same as the one for all measurements shown above.
The intermediate values for $\Delta t$ show a gradual transition between the two extreme cases.
Here, more particles remain inside the straight guide when it is switched off for shorter times.
These only have to continue the remaining distance once the latter is switched back on.

In the second type of experiment the detection time $t_{det}$ was fixed at 4.5 ms and 7.5 ms after switching on the curved guide, and $\Delta t$ was scanned.
The results are shown in the inset of fig.\ \ref{decay}: they directly show the decay of the particle density inside the straight guide.
The dashed (solid) lines represent the data for $t_{det}$=4.5 ms (7.5 ms).
Both curves can be fitted by an exponential decay.
The resulting half-times for the decay are $\tau_{1/2}=673(18)$ $\mu$s for $t_{det}$=7.5 ms, and  $\tau_{1/2}=444(8)$ $\mu$s for $t_{det}$=4.5 ms.
It is tempting to ascribe this difference to a difference in transverse velocities for the different longitudinal velocity components probed at 4.5 and 7.5 ms.
Assuming a cylindrically symmetric distribution, half of the particles have to fly 4 mm to leave the guide transversely.
Converting $\tau_{1/2}$ would then provide a value for an average transverse velocity of 9 m/s and 6 m/s for $t_{det}$= 4.5 ms and 7.5 ms, respectively.
But since the initial distribution is a thermalized Maxwell-Boltzmann distribution such an assumption would be unreasonable.
However, a discrimination arises from the fact that molecules with high longitudinal velocity have a higher probability to get lost not by moving out of the guide sideways but longitudinally.
At $t_{det}$=4.5 ms the slowest molecule that is detected with $\Delta t$=0 has $v_z\approx$110 m/s.
After $\Delta t$=1 ms, these particles will have moved forward by 11 cm, independently of their transverse velocity.
Any molecule that is closer to the end of the straight guide than 11 cm can be considered lost after this delay.
In contrast at $t_{det}$=7.5 ms, the slowest detectable molecules at $\Delta t$=0 has $v_z\approx$70 m/s.
This lower threshold means that, at a given value of $\Delta t$, the molecules have a smaller chance to leave longitudinally, thus reducing the loss rate.

In the simulations shown in fig. \ref{switched} it is implicitly assumed that all particles are lost immediately.
The calculations can, however, reproduce the experimental data much better if a pulse with a duration 100 $\mu$s longer is assumed. 
This corresponds to the time-scale expected from the purging dynamics: the extra 100 $\mu$s would describe an averaged situation for the non-zero purging time.

\subsection{Rovibrational state population of the exit beam}
\begin{figure}
\includegraphics[width=3in]{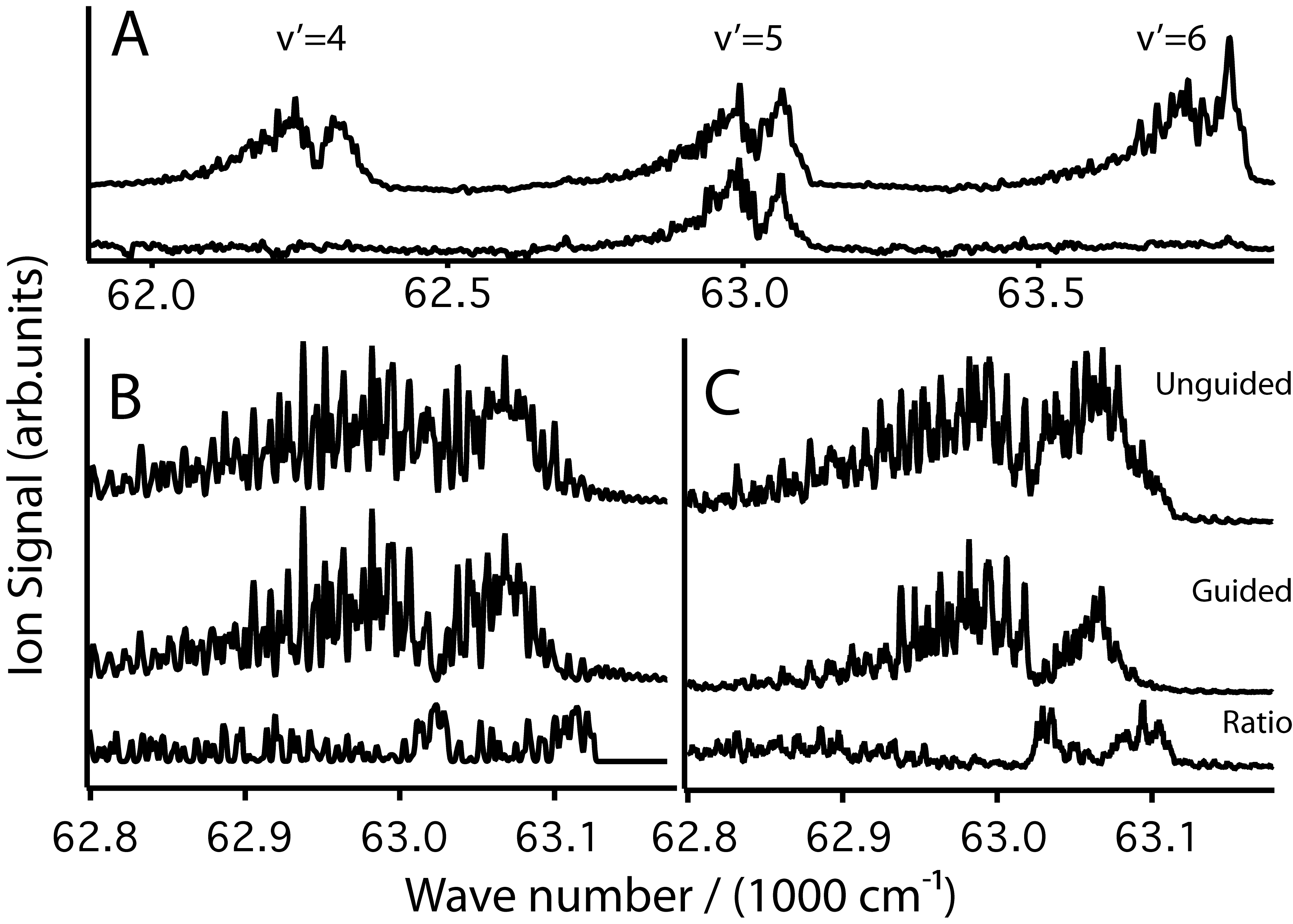}
\caption{\label{laser}A) REMPI Spectra showing three vibrational bands of the $\tilde{B}\leftarrow\tilde{X}$ transition with the guide off (upper trace) and on (lower trace). B) and C) Close-up of the $(v_2'=5)$ band without guide (top traces) and with guide (middle traces). For each case there is an simulated (panel B) and a experimental (panel C) result. Bottom traces: ratio guided/unguided.}
 \end{figure}
The exit-molecular beam is characterized by first studying the vibrational state population.
REMPI reveals the state-specific guiding-probability through the comparison of back\-ground spectra with those of guided molecules.
Figure \ref{laser}A) shows spectra of three vibrational bands of the $\tilde{B}\leftarrow\tilde{X}$ transition, recorded for background gas (upper trace) and guided molecules (bottom trace).
The vibrational excitation of the excited state is indicated above each band.
The background spectrum shows the complete progression of vibrational bands with $v_2'=4-6$, but the spectrum of the guided molecules only shows a single band with $v_2'=5$.
The reason are the symmetries of ground and excited states:
Even (odd) vibrational levels in the $v_2$ progression have $E''$ ($E'$) symmetry and can only be excited from the $\tilde{X}(0)$ ($\tilde{X}(1)$) level of the ground state.
But since the $\tilde{X}(0)$ is the lower component of the inversion doublet, all rotational levels of this state are hfs, while all rotational levels of the $\tilde{X}(1)$ state are lfs.
The electrostatic guide only guides lfs states, and all $\tilde{X}(0)$ levels are completely lost from the sample by the end of the curved guide.

Looking closer into the structure of a single vibronic band should reveal the different guiding probabilities for different rotational states.
Because the Stark effect depends on the effective dipole moment $\mu_{eff}\propto\frac{\M\K}{\J(\J+1)}$, the shape of a single vibrational band is expected to differ between background and guided molecules.
This was measured (and calculated) by again comparing the spectra of guided and unguided molecules, but at higher resolution, as shown in figs. \ref{laser}B) (simulations) and C) (experiment).
Here, the top traces show the spectra for the $\tilde{B}(v_2'=5)\leftarrow\tilde{X}(1)$ transition at a temperature of 300 K under normal conditions.
At this temperature, the spectral density is to high to allow for a complete resolution and assignment of all transitions.
Nevertheless, the comparison with the spectrum of guided molecules (middle traces in fig. \ref{laser}B) and C) shows clear differences.
The spectrum was simulated roughly by ignoring all transitions from levels with K$\leq1$.
The bottom two traces show the ratios between the simulated spectra and between the calculated spectra.
Good agreement between these two traces exists, indicating that the main change between the two situations results from the missing transitions from $K=0$ and 1 levels which have no or very low Stark effect ($\kappa\approx0$).

\begin{figure}
\includegraphics[width=3in]{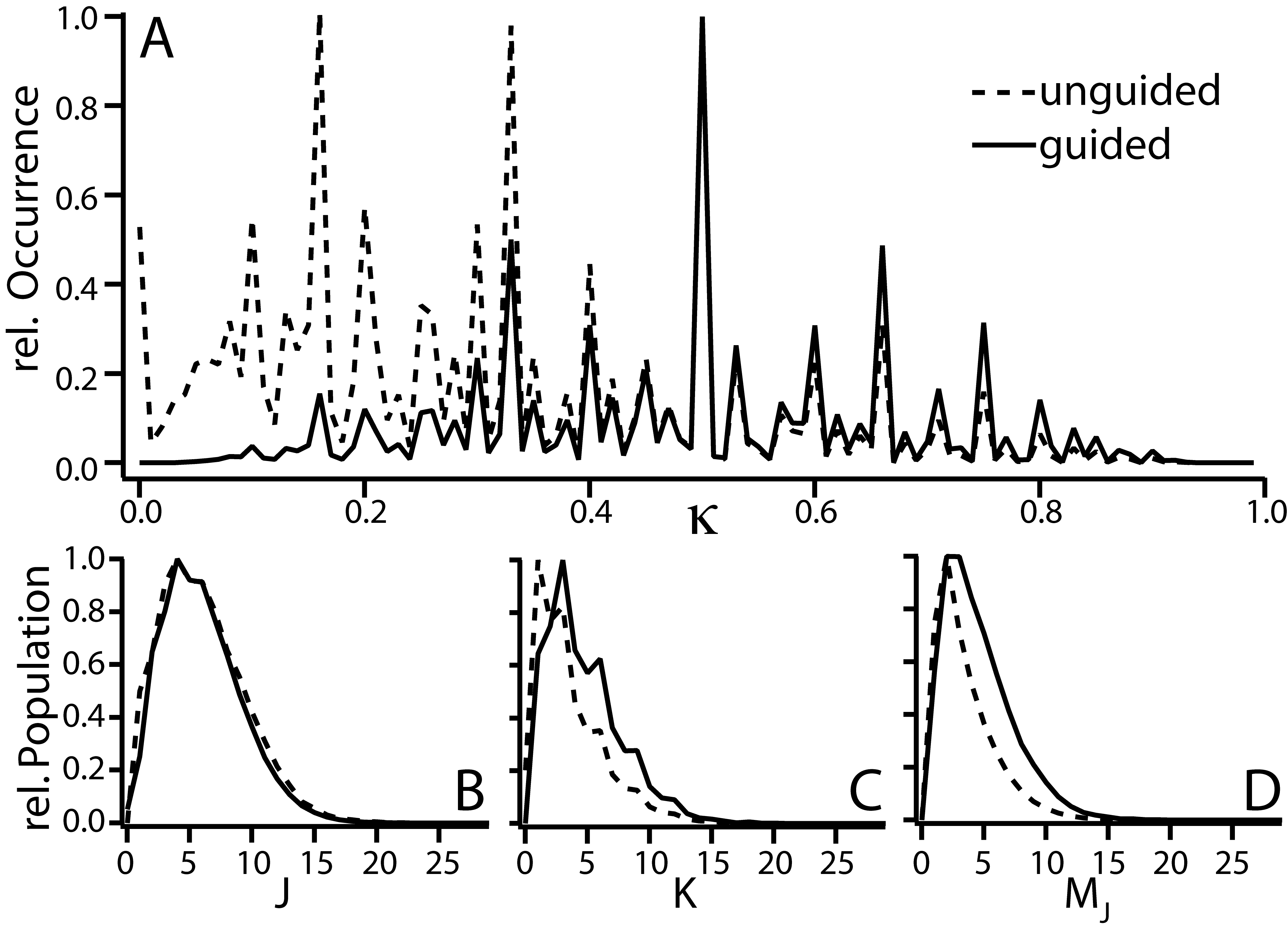}
\caption{\label{rot_states}Comparison of the calculated rotational state populations before and behind the guide. A) shows $\kappa=\frac{\M\K}{\J(\J+1)}$, being proportional to $\mu_{eff}$ before (dashed line with crosses) and after (solid line with circles) the filtering. B), C), and D): relative population for states with different values of their J--, K--, and \Mt--quantum numbers, respectively. In each panel, the dashed line is before and the solid line is after the filter.}
 \end{figure}
 Further information is extracted from trajectory simulations: Fig. \ref{rot_states}A) shows the distribution of $\kappa$-values before (dashed line) and after (solid line) the guide (normalized to the value at $\kappa=0.5$).
 The dependence of the effective dipole moment on $\kappa$ is clearly reflected in the elimination of all low values of $\kappa$ during the guiding process.
 Further understanding is provided by investigating different J, K, and \Mt-values, respectively, before and after the guide.
 Panels B)-D) of fig. \ref{rot_states} show the relative populations for states with different J (panel B), K (panel C), and \Mts (panel D).
 In each case, the dashed line with crosses (solid line with circles) shows the values before (after) the guide.
 The most striking observation is that the relative population of different J-levels remains practically unchanged.
 Larger differences are observed in K.
 Here, however, the main change is that states with low K-values are almost completely eliminated while the higher ones are all still populated.
 The main reason for the observation in panel A comes indeed from different \Mts values.
 In an electric field, all $\langle \J ,\K|$ levels are split into Stark states with potentially very low \Mt, and these states are not guided independently of the other quantum numbers.
 For a given value of J the rotational states are evenly distributed over $0<\kappa<1$, and the state guiding probability ultimately depends on the ratio between K and J:
 For any sufficiently high $\langle \J, \K|$ level the average effective dipole moment is proportional to 
 \begin{eqnarray}
 \mu_{eff} & \propto & \frac{\M\K}{\J(\J+1)} \\
  & \propto & \frac{\K}{\J},
  \end{eqnarray}
since the average value of $\bar{|\mathrm{M}_\mathrm{J}|}$ of the lfs states is $\J/2$.
The essential component is a high K-value, and the best-guided levels are those with K$\approx$J where $\kappa$ approaches 1.
Molecules with small rotational constants have rotational spectra that are sufficiently close to continuous that no significant difference is observed in the populations before and after the guide.
Because in the present setup a room-temperature effusive source was employed, the rotational temperature before the guide was 300 K, and while it is difficult to extract an actual temperature from the spectra shown here, the safest assumption is that it is still almost 300 K.

An estimate of the flux in this beam is possible by comparing the REMPI signal from the guided molecules to the signal from background-molecules: During operation, the pressure in the detection chamber rises from $<$5$\cdot$10$^{-9}$ mbar to around 2$\cdot$10$^{-8}$ mbar.
The background gas produces a signal which, under operating conditions without discrimination, is comparable in intensity to the signal from the guided molecules.
While the background pressure, which is read off of a pressure gauge some 30 cm away from the interaction region, has to be taken as an approximate value only it still allows the estimation of a density in the beam.
In the present case a density of 10$^8$ cm$^{-3}$ is assumed.
REMPI measures density and not flux, and this number can be transformed into flux by assuming a laser diameter at the focus of 0.1 mm.
Taking an average velocity of 120 m/s, this corresponds to a flux of $\approx$10$^9$/s.
Note that even though the laser can only be set to detect a single, or very few, states at a time, these numbers still correspond to the total density and flux, because the comparison with the background also only includes a single transition.
While it carries a considerable error, the so estimated flux here is comparable to results from simulations as well as the values reported by other groups \cite{Sommer:2009p3065}.

\section{Conclusion}
We have demonstrated the velocity filtering of deuterated ammonia in an electrostatic hexapole guide with a large radius of curvature and combined it with time- and state-selective detection of the filtered molecules using REMPI.
Continuous molecular beams with translational temperatures below 10 K were generated.
The translational temperature could be further reduced by using a segmented guide with a straight and a curved segment.
Switching the high voltage on the straight guide segment enabled the discrimination against slow molecules and thus a further reduction of the translational temperature.
The combination of different switching times with different voltages on the curved guide segment allows the production of a molecular sample with a selectable temperature and at a selectable mean velocity in the laboratory frame-of-reference.

REMPI detection enabled the characterization of the rotational temperature of a velocity-filtered sample.
It showed that the rotational state distribution changes only marginally between before and after the guide, with the main difference being the elimination of all levels with $\K\leq1$.
In contrast, complete purging of hfs was shown by comparing the intensities of vibrational bands with even and odd $v_2$ quantum numbers.
For the preparation of translationally and internally cold molecules by this technique, the source would have to be replaced by a method to produce internally cold molecules.
It has been shown previously that buffer gas cooling is a viable method to achieve this goal \cite{Sommer:2009p3065,Patterson:2007p4868}.
Alternatively, a rotating nozzle producing a continuous supersonic expansion would be an attractive alternative to the pure effusive source used here.\cite{Gupta:2001p4871}

We are planning to make use of another attractive feature of the hexapole geometry of the guide: it can also be used to generate a dipole field, e.g., by applying a strong positive and a strong negative voltage to the two outermost electrodes of the guide, and low voltage to the innermost ones \cite{Crompvoets:2005p4878}. 
This generates a dipole filed pointing to the inside of the curvature, and the molecules are not confined from the inside of the guide; only the centrifugal force keeps them in the guide.
Consequently, using a particular electric field selects a velocity-band from the initial distribution:
Fast molecules are ejected outwards because the confining force is insufficient while slow molecules are pushed inwards because the centrifugal force is insufficient.
A particularly interesting aspect of this configuration is that it can relatively easily be extended to hfs states, simply by reversing the direction of the dipole.
Combining a dipole guide with the independent guide segments furthermore enables the decoupling of velocity- and state-selection: 
The selection by segment-switching is based purely on the forward-velocity, and a curved guide can then be used for state-selection.

\section{Acknowledgments}
We would like to thank Gerard Meijer (Fritz-Haber-Institute, Berlin) for providing parts of the equipment and for continuous support throughout the project.
We also acknowledge help by Cynthia Heiner (Fritz-Haber-Institute, Berlin) and Maxwell Parsons (Harvard University, Cambridge(USA)) during different stages of this project. 
This work is funded by the Swiss National Science Foundation.


\end{document}